\documentclass[12pt]{article}
\usepackage{graphicx}
\usepackage[small]{titlesec}
\usepackage[utf8]{inputenc} 
\usepackage{amsmath}
\usepackage[T1]{fontenc}
\usepackage[english]{babel}
\usepackage[left=2cm,right=2cm,top=2cm,bottom=2cm,bindingoffset=0cm]{geometry}

\graphicspath{{figures/}}

\usepackage{setspace}
\usepackage{subcaption}
\usepackage{youngtab}
\usepackage{epigraph}
\usepackage{upgreek}

\usepackage[labelfont=bf, margin=20pt,font={small}]{caption}

\usepackage{amsmath}
\usepackage{esint}
\usepackage{bm}
\usepackage{
	tikz,
	amsmath,
	amssymb,
	amsfonts,
	amsthm,
	amscd,
	comment,
	epsfig,
	color,
	setspace,
	bbold,
	verbatim,
	slashed
}
\usepackage{upgreek}
\usepackage{tikz}
\usetikzlibrary{arrows}
\usepackage{mathrsfs}
\usepackage{multirow}
\usepackage{simpler-wick}
\usepackage{cite}
\usepackage{tensor}

\newcommand\beq{\begin{equation}}
	\newcommand\eeq{\end{equation}}

\tikzset{cross/.style={cross out, draw=black, minimum size=2*(#1-\pgflinewidth), inner sep=0pt, outer sep=0pt},
	cross/.default={5pt}}
\usepackage{mathtools}
\usepackage{graphicx}
\usepackage{authblk}
\usepackage{xcolor}
\usepackage{bbold}
\usepackage[makeroom]{cancel}
\usepackage[hidelinks]{hyperref}

\definecolor{linkcolor}{rgb}{0,0,1}
\definecolor{urlcolor}{rgb}{0,0,1}

\hypersetup{pdfstartview=FitH, linkcolor=linkcolor,urlcolor=urlcolor,citecolor=urlcolor, colorlinks=true}
\author[1]{V.~I.~Lapushkin\thanks{\href{mailto:lapushkin.vi@phystech.edu}{lapushkin.vi@phystech.edu}}}
\affil[1]{\itshape Institutskii per, 9, Moscow Institute of Physics and Technology, 141700, Dolgoprudny, Russia}
\title{\textcolor{black}{Simplification of nonlinear equations for a field operator}}

\begin{document}
	\date{}
	\maketitle
	
	\begin{abstract}
		In this paper, we study different properties of the motion  equations of interacting fields. In the second section, we prove that "Wightman's" fields (we use only a subset of Wightman's axioms) are unitarily equivalent to some operators on the vector space ${\cal F}$ (with one mathematical assumption). In the third section, we introduce \(L^{\infty}\) and  \(DL\) Hilbert spaces, which are convenient for analyzing field equations, particularly the equations for \(\phi^3\) theory. Remarkably, we have managed to reduce the equation of motion for \(\phi^3\) to a quadratic matrix equation with matrices over a separable Hilbert space in the fourth section. Also, in the appendix, we have done the same for QCD. Furthermore, we prove the existence of solution to the motion equations of one toy model non-renormalizable theory in the fifth section.  
	\end{abstract}
	
	\tableofcontents
	\newpage
	
	\section{Introduction}
	In the Wightman axioms [1] and other works on axiomatic quantum field theory, the properties of interacting fields and the Hilbert space on which these fields are defined were extensively studied. However, a convenient form of the Hilbert space where these fields actually operate was never presented. In works where this was mentioned [2], more attention was paid to the properties of this Hilbert space (denoted in this work as \( {\cal F} \)) than to the properties of the field equations over this space.  
	
	In this paper, we study various properties of the equations of motion for interacting fields. In the second section, we prove that "Wightman's" fields (using only a subset of Wightman's axioms) are unitarily equivalent to some operators on the space ${\cal F}$ (with one mathematical assumption). In the third section, we introduce \(L^{\infty}\) and \(DL\) Hilbert spaces, convenient for analyzing field equations, particularly for \(\phi^3\) theory. Remarkably, we have reduced the equation of motion for \(\phi^3\) to a quadratic matrix equation with matrices over a separable Hilbert space in the fourth section. Also, in the appendix, we have done the same for QCD. Furthermore, we prove the existence of solution to the motion equations of one toy model non-renormalizable theory in the fifth section.  
	
	\section{Modification of the Wightman Reconstruction Theorem}
	Usually, the GNS construction is used for C*-algebras, but here we will need to use similar constructions for topological *-algebras. We will call this the "GNS" construction. First, we will prove the "GNS" construction and then modify certain steps in the proof to derive a result more suitable for further applications.
	
	\textbf{"GNS" Construction}
	Suppose we have a positive normalized functional \( F \) on a *-algebra \( \mathcal{U} \). In this case, we can construct a Hilbert space \( \mathcal{H}_{F} \) on which a cyclic representation of the *-algebra is realized with a cyclic unit vector \( \Phi_{F} \), and the following formula holds:
	\[F(A)=\langle \Phi_{F}, \pi_{F}(A)\Phi_{F}\rangle\]
	
	\textbf{Proof}: Consider the subset \( \mathcal{J} \) of the topological *-algebra \( \mathcal{U} \):
	\[\mathcal{J}=\{A\in \mathcal{U}: F(A^{\dagger}A)=0\}\]
	From the properties of positive functionals, it follows that this set forms a *-algebra, specifically a left ideal in \( \mathcal{U} \). Now, let \( \mathcal{H}_{F} \) denote the quotient space \( \mathcal{U}/\mathcal{J} \). Define a mapping from \( \mathcal{U} \) to \( \mathcal{H}_{F} \):
	\[\xi_{F}(A)\in \mathcal{H}_{F}\]
	Where $\xi_F$ is quotient map. That is, for any \( X \in \mathcal{J} \), \( \xi_F(X)=0 \). The representation \( \pi_F \) is defined as follows:
	\[\pi_F(A)\xi_F(B)=\xi_F(AB)\]
	This definition is correct because the new element depends on \( B \) through \( \xi_F(B) \), since for any \( X \in \mathcal{J} \):
	\[\xi_F(A(B+X))=\xi_F(AB+AX)=\pi_F(A)\xi_F(B+X)=\pi_F(A)\xi_F(B)=\xi_F(AB)\]
	Introducing the unit vector \( \Phi_F=\xi_F(E) \), where \( E \) is the unit element of the algebra, completes the construction. This vector is normalized due to the normalization of the functional \( F \). $\square$
	
	Suppose there exists an automorphism \( \gamma \) of the *-algebra \( \mathcal{U} \). Moreover, the positive normalized functional \( F \) is invariant under this automorphism, i.e.:
	\[F(\gamma(A))=F(A)\]
	Then, we can define a unitary operator \( U_F \):
	\[U_F\xi_F(A)=\xi_F(\gamma(A))\]
	\[\langle \xi_F(A),\xi_F(B)\rangle=F(A^{\dagger}B)=F(\gamma(A)^{\dagger}\gamma(B))=\langle U_F\xi_F(A),U_F\xi_F(B)\rangle\]
	\[U_F\Phi_F=\Phi_F\]
	The last equality holds because \( \gamma(E)=E \). Thus, the unitary operator corresponding to the automorphism \( \gamma \) acts on \( \mathcal{H}_{F} \), and the vector \( \Phi_F \) is invariant under this transformation.
	
	\textbf{Lemma on Quotient Space}: Suppose there is a vector space \( V_1 \), a set of automorphisms \( \gamma_{\lambda} \) forming a group, and a degenerate scalar product defined by a bilinear form \( \Omega \) that is invariant under automorphisms. Define \( V_2=\{\psi\in V_1: \Omega(\psi,\psi)=0\} \). Then, we can construct an isomorphism between the quotient space \( V_1/V_2 \) and some subspace \( V \) in \( V_1 \), and define a Hilbert scalar product and a set of automorphisms \( \gamma_{\lambda} \) on this subspace.
	
	\textbf{Proof}: Suppose there exists an isomorphism between \( A\subset V_1/V_2 \) and a subset \( V_A \) in \( V_1 \). Take some vector \( \psi \) in \( V_1/V_2 \) not lying in \( A \). Associate with it an element \( \Psi \) from \( V_1 \), which under the homomorphism from \( V_1 \) to \( V_1/V_2 \) maps to \( \psi \). Also, associate with each vector of the form \( \gamma_{\lambda}(\psi) \) the vector \( \gamma_{\lambda}(\Psi) \). Additionally, the zero vector from \( A \) should be associated with the zero vector from \( V_1/V_2 \). This mapping defines the required isomorphism but now for a larger space (the linear span of \( A \), \( \gamma_{\lambda}(\psi) \) and the linear span of \( V_A \), \( \gamma_{\lambda}(\Psi) \)). After applying Zorn's lemma, we obtain the existence of an isomorphism between \( V \) and \( V_1/V_2 \). Define the scalar product on \( V \) via \( \Omega \). Moreover, the scalar product defined in this way is non-degenerate. Otherwise, there would exist a vector \( \psi\in V \) such that it belongs to \( V_2 \), but then this vector under the homomorphism from \( V_1 \) to \( V_1/V_2 \) would map to the zero vector, and hence, by the construction of the isomorphism, it would be zero in \( V \). The closure of \( V \) under the action of the group of automorphisms follows directly from the construction of this space. $\square$
	
	From the construction of the isomorphism between the spaces \( V \) and \( V_1/V_2 \), it is clear that \( V_1=V_2\oplus V \).
	
	\subsection{"GNS" Construction in the Context of Wightman Functionals}
	Let there be "Wightman's" fields \( \hat{\phi}^{\varkappa}(x) \), where \( \varkappa \) indicates a type of particle. By "Wightman's" fields, we mean that these objects are defined similarly to Wightman's fields but do not satisfy two of the seven Wightman axioms: Positivity of spectrum $\hat{P}_{\mu}$ and Causality. They are not necessary for the construction of the Hilbert space; moreover, Positiveness of spectrum does not work in $\phi^3$ [3]. Each field is a tensor or spin-tensor quantity with a finite number of components \( \hat{\phi}^{\varkappa}_l(x) \) ($x$-is d dimensional vector, but usually we will write \( x \) without an index). We can also define Hermitian conjugation as \( \hat{\phi}^{\varkappa\dagger}_l(x)=\hat{\phi}^{\overline{\varkappa}}_{\overline{l}}(x) \). Now we will recall part of the Wightman reconstruction theorem, but as before, we will not use two of the seven Wightman axioms ([4] 8.2.A). First, we need to construct the space \( {\cal F} \):
	\[{\cal F}=\{(f_0,f^{(\varkappa_1)}_{l_1}(x_1),...);f\in C; f^{(\varkappa_1...\varkappa_n)}_{l_1...l_n}(x_1,...,x_n)\in S(R^{dn})\}\]
	On it, we can define multiplication and involution operations:
	\[(f\otimes g)^{(\varkappa_1...\varkappa_n)}_{l_1...l_n}(x_1,...,x_n)=\sum_{k=0}^{k=n}f^{(\varkappa_1...\varkappa_k)}_{l_1...l_k}(x_1,...,x_k)g^{(\varkappa_{k+1}...\varkappa_{n})}_{l_{k+1}...l_n}(x_{k+1},...,x_n)\]
	\[(f^{\dagger})^{(\varkappa_1...\varkappa_n)}_{l_1...l_n}(x_1,...,x_n)=f^{(\overline{\varkappa}_n...\overline{\varkappa}_1)*}_{\overline{l}_n...\overline{l}_1}(x_n,...,x_1)\]
	We can also introduce a unit element \( E=(1,0,...) \) and define a representation of the Poincaré group:
	\[(\hat{U}(a,\Lambda(\underset{\sim}{\Lambda}))f)^{(\varkappa_1...\varkappa_n)}_{l_1...l_n}(x_1,...,x_n)=\sum_{m_1,..,m_n}V^{(\varkappa_1)}_{l_1m_1}(\underset{\sim}{\Lambda})...V^{(\varkappa_n)}_{l_nm_n}(\underset{\sim}{\Lambda})f^{(\varkappa_1...\varkappa_n)}_{m_1...m_n}(\Lambda(\underset{\sim}{\Lambda})^{-1}(x_1-a),...,\Lambda(\underset{\sim}{\Lambda})^{-1}(x_n-a))\]
	Where \( \underset{\sim}{\Lambda}\in SL(2,C) \) and \( V^{(\varkappa)}_{lm}(\underset{\sim}{\Lambda}) \) is a real or complex finite-dimensional representation of \( SL(2,C) \). This holds if \( d=4 \); otherwise, we need to take the universal covering group of the Lorentz group instead of \( SL(2,C) \). Now, we can consider \( {\cal F} \) as a topological *-algebra. Define the "Wightman's" functional as follows:
	\[W(f)=w_0f_0+\sum^{\infty}_{n=1}\sum_{\varkappa_1,...,\varkappa_n,l_1,...,l_n}\int...\int w^{(\varkappa_1...\varkappa_n)}_{l_1...l_n}(x_1,...,x_n)f^{(\varkappa_1...\varkappa_n)}_{l_1...l_n}(x_1,...,x_n)dx_1...dx_n\]
	Where \( w^{[n]}(x_1,...,x_n) \) is the n-point "Wightman's" function. From the five remaining Wightman axioms, it follows that this functional is positive, normalized, and Poincaré-invariant. Therefore, we can apply the "GNS" construction to this functional and obtain the Hilbert space \( \mathcal{H}={\cal F}/J \), where \( J \) is the set of elements annihilated by the functional.
	
	From the lemma on the quotient space, it follows that \( {\cal F}=\mathcal{H}\oplus J \), and on \( \mathcal{H} \), the scalar product is defined as:
	\[(f,g)_W=W(f^{\dagger}\otimes g)\]
	The scalar product between vectors from \( \mathcal{H} \) and \( J \) is defined as 0. The scalar product in \( J \) is given by:
	\[(f,g)_W=f_0^*g_0+\sum^{\infty}_{n=1}\sum_{\varkappa_1,...,\varkappa_n,l_1,...,l_n}\int...\int f^{(\varkappa_1...\varkappa_n)*}_{l_1...l_n}(x_1,...,x_n)g^{(\varkappa_1...\varkappa_n)}_{l_1...l_n}(x_1,...,x_n)dx_1...dx_n\]
	But som of the vectors from ${\cal J}$ would have the infinity norm in this case. We can take ${\cal J'}$- dense set of vectors with finite norm in ${\cal J}$(${\cal F'}={\cal J'}\oplus{\cal H}$- dense set of vectors with finite norm in ${\cal F}$).Thus, we have defined the Wightman scalar product on the vector space \( {\cal F'} \), and this scalar product is Poincaré-invariant due to the closure of the subspaces under the Poincaré group, and Poincaré-invariance of two scalar products on each of subspces. From the Wightman reconstruction theorem, it follows that all fields \( \varkappa \) in the theory can be represented by operator-valued genralized functions over the space \( \mathcal{H} \). Moreover, the vacuum vector \( |0\rangle=(1,0,0....) \). The action of the field operators is defined on \( \mathcal{H} \) or on the subspace \( \mathcal{H} \) in \({\cal F'}\) ([4] 8.3.B). We can also define the action of the fields on \( J' \), but this space and this action do not affect the physics. Therefore, this action can be defined arbitrarily, but to ensure consistency, we will define fields on \( J' \) as zero operators. Thus, in solving the equations, it must be remembered that they are valid only on some subspace of ${\cal F}$. We can even introduce \( \hat{\Pi} \), the projector onto \( \mathcal{H} \), and henceforth, we will write the equations with this projector.
	\subsection{Mathematical assumption}
	Suppose on a vector space \( V \), there are two different scalar products: \( \langle.,.\rangle_1 \) and \( \langle.,.\rangle_2 \). We will assume that in the proposed construction, there exists an automorphism \( \gamma \) such that for all \( \psi_1,\psi_2\in V \):
	\[\langle \psi_1,\psi_2\rangle_1=\langle \gamma(\psi_1),\gamma(\psi_2)\rangle_2\]
	Then, we can diagonalize the "Wightman's" functional (with this assumption, \( (.,.)_W=\langle.,.\rangle_1 \) and \(\langle.,.\rangle=\langle.,.\rangle_2 \)-is naturally defined scalar product like $(.,.)_W$ on J). Thus, we can define all fields on ${\cal F}$ with the normal scalar product \( \langle.,.\rangle \). In fact, a more basic assumption is enough for us, that we can choose such $\langle.,.\rangle_2$ that $|0\rangle$ is orthogonal to all vectors with $f_0=0$ and that it's norm is 1. 
	
	\section{$L^{\infty}$ and $DL$}
	The most important operator appearing in the motion equations is $\hat{P}_{\mu}$-generator of translations($\hat{U}(a,1)=e^{-i(\hat{P}a)}$). So we need to find some space with a simple spectrum of the operator $\hat{P}_{\mu}$ and so that there is a simple bijection between this space and ${\cal F}$. First, we need to find an operator $\hat{V}_k:S(R^{kd})\rightarrow L_0(R^{d})$ such that:
	\[\hat{P}_{\mu}f(x_1,...,x_k)=i(\frac{\partial}{\partial x^{\mu}_1}+...+\frac{\partial}{\partial x^{\mu}_k})f(x_1,...,x_k)\]
	\[\hat{P}_{\mu}(\hat{V}_nf)(x)=\hat{P}_{\mu}f_V(x)=i\frac{\partial}{\partial x^{\mu}}f_V(x)\]
	Where $f(x_1,...,x_k)\in S(R^{kd})$, $(\hat{V}_{k}f)(x)=f_V(x)\in L_0(R^d)$. The most convenient form of this operator would be in the basis $e_{n_1,...,n_k}(x_1,...,x_n)=\prod_{\mu}((\frac{\partial}{\partial x^{\mu}_1})^{n^{\mu}_1}...(\frac{\partial}{\partial x^{\mu}_k})^{n^{\mu}_k}e^{-\frac{(x^{\mu}_1)^2+...+(x^{\mu}_k)^2}{2}})$. We can decompose $f(x_1,...,x_n)$ into this basis:
	\[f(x_1,...,x_n)=\sum_{n_1,...,n_k}f_{n_1,...,n_k}e_{n_1,...,n_k}(x_1,...,x_n)\]
	Our purpose is to find a basis $\hat{V}_n(e_{n_1,...,n_k}(x_1,...,x_n))=e_{n_1,...,n_k}(x)$ such that:
	\[\frac{\partial}{\partial x^{\mu}}e_{n_1,...,n_k}(x)=e_{n_1+e(\mu),n_2,...,n_k}(x)+e_{n_1,n_2+e(\mu),...,n_k}(x)+...+e_{n_1,n_2,...,n_k+e(\mu)}(x)\]
	Where $e(\mu)$ is a d-vector with only one non-zero component $(e(\mu))^{\nu}=\delta_{\mu,\nu}$. It would be more convenient to find the Fourier transformation of this basis:
	\[-ip^{\mu}\tilde{e}_{n_1,...,n_k}(p)=\tilde{e}_{n_1+e(\mu),n_2,...,n_k}(p)+\tilde{e}_{n_1,n_2+e(\mu),...,n_k}(p)+...+\tilde{e}_{n_1,n_2,...,n_k+e(\mu)}(p)\]
	We can build this basis using the function $\pi(n_1,...,n_{k-1})$-a bijection between $\mathbb{N}^{k-1}$ and $\mathbb{Z}$:
	\[\tilde{e}_{n_1,...,n_{k-1},1}(p)=I(p^{0}\in[\pi(n_1,...,n_{k-1});\pi(n_1,...,n_{k-1})+1))\]
	All other functions for different $n_k$ can be constructed using recurrence relations. So we build the new space with a new scalar product and new spectrum of the operator $\hat{P}_{\mu}$. Now we get a convenient representation of the space ${\cal F}$: $L^{\infty}=\{|f\rangle=(f_0,f^{(\varkappa_1)}_{l_1}(p),f^{(\varkappa_1\varkappa_2)}_{l_1l_2}(p),...);f_0\in C, f^{(\varkappa_1...\varkappa_n)}_{l_1...l_n}(p)\in L_0(R^d)\}$\footnote{Of course, in fact, the indicated bijection is carried out between vectors from ${\cal F}$ with finite norm and from $L^{\infty}$ with finite norm, in fact, the condition on functions from $L^{\infty}$ is more complex than the fact that they are simply measurable, but we will not prescribe this condition, since it is very cumbersome and obvious (the condition on the finiteness of the norm in the new space)}. An exact bijection ${\cal F}\rightarrow L^{\infty}$ can be built using $\hat{V}_n$:
	\[(f_0,f^{(\varkappa_1)}_{l_1}(x_1),f^{(\varkappa_1\varkappa_2)}_{l_1l_2}(x_1,x_2),...)\rightarrow(f_0,f^{(\varkappa_1)}_{l_1}(x),(\hat{V}_2f^{(\varkappa_1\varkappa_2)}_{l_1l_2})(x),...)\rightarrow(f_0,f^{(\varkappa_1)}_{l_1}(p),f^{(\varkappa_1\varkappa_2)}_{l_1l_2}(p),...)\]
	The last bijection is Fourier transformation. The representation of the translation group is given by the translation generator representation:
	\[\hat{P}^{\mu}(f_0,f^{(\varkappa_1)}_{l_1}(p),f^{(\varkappa_1\varkappa_2)}_{l_1l_2}(p),...)=(0,p^{\mu}f^{(\varkappa_1)}_{l_1}(p),p^{\mu}f^{(\varkappa_1\varkappa_2)}_{l_1l_2}(p),...)\]
	The last one step in our construction is the representation of the Lorentz group. First of all we will write the Lorentz transormations for ${\cal F}$:
	\[(\hat{U}(1,\Lambda(\underset{\sim}{\Lambda}))f)^{(\varkappa_1...\varkappa_n)}_{l_1...l_n}(x_1,...,x_n)=\sum_{m_1,..,m_n}V^{(\varkappa_1)}_{l_1m_1}(\underset{\sim}{\Lambda})...V^{(\varkappa_n)}_{l_nm_n}(\underset{\sim}{\Lambda})f^{(\varkappa_1...\varkappa_n)}_{m_1...m_n}(\Lambda(\underset{\sim}{\Lambda})^{-1}x_1,...,\Lambda(\underset{\sim}{\Lambda})^{-1}x_n)\]
	We can introduce new transformation $\hat{U}_0(\Lambda(\underset{\sim}{\Lambda}))$:
	\[(\hat{U}_0(\Lambda(\underset{\sim}{\Lambda}))f)^{(\varkappa_1...\varkappa_n)}_{l_1...l_n}(p^{\mu})=\sum_{m_1,..,m_n}V^{(\varkappa_1)}_{l_1m_1}(\underset{\sim}{\Lambda})...V^{(\varkappa_n)}_{l_nm_n}(\underset{\sim}{\Lambda})f^{(\varkappa_1...\varkappa_n)}_{m_1...m_n}(\Lambda^{\mu}_{\nu}(\underset{\sim}{\Lambda})p^{\nu})\]
	Now we can use two obvious properties for $\hat{U}_0(\Lambda(\underset{\sim}{\Lambda}))$ and $\hat{U}(1,\Lambda(\underset{\sim}{\Lambda}))$: \[\hat{U}_0(\Lambda(\underset{\sim}{\Lambda}))\hat{P}^{\mu}\hat{U}^{-1}_0(\Lambda(\underset{\sim}{\Lambda}))=\Lambda^{\mu}_{\nu}(\underset{\sim}{\Lambda})\hat{P}^{\nu}=\hat{U}(1,\Lambda(\underset{\sim}{\Lambda}))\hat{P}^{\mu}\hat{U}^{-1}(1,\Lambda(\underset{\sim}{\Lambda}))\]
	Where we ment under $\hat{U}(1,\Lambda(\underset{\sim}{\Lambda}))$ the acting of this operator on the spsce $L^{\infty}$.We can introduce projectors $\hat{P}^{(\varkappa'_1,...,\varkappa'_n)}_{l'_1,...,l'_n}(f_0,f^{(\varkappa_1)}_{l_1}(p),...)=(0,...,0,\delta_{\varkappa_1,\varkappa'_1}...\delta_{l_1,l'_1}....f^{(\varkappa_1,...,\varkappa_n)}_{l_1,...,l_n}(p),0,...)$. So we now obtain that $\hat{U}^{-1}_0(\Lambda(\underset{\sim}{\Lambda}))\hat{U}(1,\Lambda(\underset{\sim}{\Lambda}))$ commutes with all $\hat{P}^{(\varkappa'_1,...,\varkappa'_n)}_{l'_1,...,l'_n}$and $\hat{P}^{\mu}$. So we can obtain the next formula:
	\[(\hat{U}(1,\Lambda(\underset{\sim}{\Lambda}))f)^{(\varkappa_1...\varkappa_n)}_{l_1...l_n}(p^{\mu})=\sum_{m_1,..,m_n}V^{(\varkappa_1)}_{l_1m_1}(\underset{\sim}{\Lambda})...V^{(\varkappa_n)}_{l_nm_n}(\underset{\sim}{\Lambda})f^{(\varkappa_1...\varkappa_n)}_{m_1...m_n}(\Lambda^{\mu}_{\nu}(\underset{\sim}{\Lambda})p^{\nu})U^{(\varkappa_1...\varkappa_n)}_{m_1...m_n}(p;\underset{\sim}{\Lambda})\]
	Where $U^{(\varkappa_1...\varkappa_n)}_{m_1...m_n}(p;\underset{\sim}{\Lambda})$ are any functions. In the case where we have only scalar fields we can introduce $KL_{E}$ in Euclidean signature (the same construction for Minkovsky's signature is way harder). First we will introduce $KL_{E}=\{(f_0,f(p)); f_0\in C, f(p)\in L(R^d)\}$ with the next exact form of the transformation $L^{\infty}\rightarrow KL_{E}$:
	\[(f_0,f^{(\varkappa_1)}(\vec{p}),f^{(\varkappa_1\varkappa_2)}(\vec{p}),...)\rightarrow(f_0,f_1(\vec{p}),f_2(\vec{p}),...)\rightarrow(f_0,\sum^{+\infty}_{k=1}f_k(ctg(\{|\vec{p}|\}\pi))I_k([|\vec{p}|]\pi=k-1))\]
	Where first one bijection is redesignation of indices, $[.]$-is whole part of the number and $\{.\}$-is a fractional part. The action of the rotatin group is determined like:
	\[\hat{U}(1,R)(f_0,f(\vec{p}))=(f_0,U(\vec{p};R)f(R\vec{p}))\]
	Where $U(\vec{p};R)$ is some function (the transformation of functions $U^{(\varkappa_1...\varkappa_n)}_{m_1...m_n}(p;\underset{\sim}{\Lambda})$). And $R$ - is any rotation. So $R\vec{p}$-rotated vector $\vec{p}$. Also there is not difficult view of the translator generator:
	\[\hat{\vec{P}}(f_0,f(\vec{p}))=(0,\frac{ctg(\{|\vec{p}|\})\vec{p}}{\pi|\vec{p}|}f(\vec{p}))\]
	Finally we can introduce one more space convinient for work (for any signature and types of particles, but we don't know the exact representation of the rotation group on it) $DL=\{(f_0,f(p)); f_0\in C, f(p)\in L(R^d)\}$. We can write the exact form of the transformation $L^{\infty}\rightarrow DL$:
	\[(f_0,f^{(\varkappa_1)}_{l_1}(p),...)\rightarrow(f_0,\sum^{+\infty}_{n=1}f_n(tg(p^{0}-\frac{\pi}{2}),...,tg(p^{d-1}-\frac{\pi}{2}))\prod_{\mu=0,..,d-1}I(p^{\mu}\in[\pi(n-1),\pi n)))\]
	Also, we can write the translation generator:
	\[\hat{P}_{\mu}(f_0,f(p))=(0,tg(p^{\mu}-\frac{\pi}{2})f(p))\]
	So now we can work with fields over simple spaces $L^{\infty}$, $KL_{E}$ and $DL$, not over ${\cal F}$. 
	\section{General Properties of Fields}
	We would like to understand the following equations:
	\[\partial_{\mu}\frac{\partial \hat{L}_0(x)}{\partial\partial_{\mu}\hat{\phi}^{(\varkappa)}_l(x)}-\frac{\partial \hat{L}_0(x)}{\partial\hat{\phi}^{(\varkappa)}_l(x)}=-N\left(\frac{\partial}{\partial \hat{\phi}^{(\varkappa)}_l(x)}V(\hat{\phi}^{(1)}_{l_1}(x),...,\hat{\phi}^{(n)}_{l_n}(x))\right)\]
	Where \( L_0 \) is the free Lagrangian and \(N\) is normally ordered product, it will be defined later; for now, we will only say that its vacuum expectation value is zero for our purposes. According to the previously described construction, we can arbitrarily extend the fields to all of \({\cal F}\) (or \( L^{\infty} \) or \(DL\)) and write them in the following form:
	\[\hat{\Pi}\left(\partial_{\mu}\frac{\partial \hat{L}_0(x)}{\partial\partial_{\mu}\hat{\phi}^{(\varkappa)}_l(x)}-\frac{\partial \hat{L}_0(x)}{\partial\hat{\phi}^{(\varkappa)}_l(x)}\right)\hat{\Pi}=-\hat{\Pi}N\left(\frac{\partial}{\partial \hat{\phi}^{(\varkappa)}_l(x)}V(\hat{\phi}^{(1)}_{l_1}(x),...,\hat{\phi}^{(n)}_{l_n}(x))\right)\hat{\Pi}\]
	Where \( \hat{\Pi} \) was introduced earlier—it is the projector onto \( \mathcal{H} \). Defining \( \mathcal{H} \) is straightforward; it is the closure of the set of vectors obtained by the powers of the field operators acting on the vacuum vector. It is also important to note that this projector commutes with \( \hat{U}(a,1) \) due to the closure of \( \mathcal{H} \) and \( J \) under Poincaré transformations (and in particular, under the translation group). The projector also commutes with \( \hat{\phi}(x) \) because the field operator maps \( \mathcal{H} \) to \( \mathcal{H} \) and \( J \) to \( J \). The  motion equations are multiplied on both sides by $\hat{\Pi}$, so if any solution of these equations exists, we can choose the solution \( \phi \) that equals zero on \( J \), because that part of operator is not involved in motion equations. So $1-\hat{\Pi}$ is the projector onto the eigenvectors of $\hat{\phi}^{(\varkappa)}_l(x)$ with zero eigenvalues ($1-\hat{\Pi}\leq I(\hat{\phi}^{(\varkappa)}_l(x)=0)$). This subspace does not depend on $x$ because $\hat{\Pi}$ commutes with $\hat{U}(a,1)$.
	\[\langle 0|\left(\partial_{\mu}\frac{\partial \hat{L}_0(x)}{\partial\partial_{\mu}\hat{\phi}^{(\varkappa)}_l(x)}-\frac{\partial \hat{L}_0(x)}{\partial\hat{\phi}^{(\varkappa)}_l(x)}\right)|0\rangle=-\langle 0|N\left(\frac{\partial}{\partial \hat{\phi}^{(\varkappa)}_l(x)}V(\hat{\phi}^{(1)}_{l_1}(x),...,\hat{\phi}^{(n)}_{l_n}(x))\right)|0\rangle=0\]
	From Poincaré covariance, it follows that:
	\[\hat{U}(a,1)\hat{\tilde{\phi}}^{(\varkappa)}_l(x)\hat{U}^{-1}(a,1)=\hat{\phi}^{(\varkappa)}_l(x+a)\]
	And from the invariance of the vacuum under Poincaré transformations, it follows that:
	\[\partial_{\mu}\langle0|\hat{\phi}^{(\varkappa)}_l(x)|0\rangle=(\partial_{a \mu}\langle0|\hat{\phi}^{(\varkappa)}_l(x+a)|0\rangle)|_{a=0}=(\partial_{a \mu}\langle0|\hat{U}(a,1)\hat{\phi}^{(\varkappa)}_l(x)\hat{U}^{-1}(a,1)|0\rangle)|_{a=0}=0\]
	We have two cases. The first is when \( \varkappa \) is a boson:
	\[0=\langle 0|\left(\partial_{\mu}\frac{\partial \hat{L}_0(x)}{\partial\partial_{\mu}\hat{\phi}^{(\varkappa)}_l(x)}-\frac{\partial \hat{L}_0(x)}{\partial\hat{\phi}^{(\varkappa)}_l(x)}\right)|0\rangle=m^2_{\varkappa}\langle0|\hat{\phi}^{(\varkappa)}_l(x)|0\rangle\]
	The second is when \( \varkappa \) is a fermion, then:
	\[0=\langle 0|\left(\partial_{\mu}\frac{\partial \hat{L}_0(x)}{\partial\partial_{\mu}\hat{\phi}^{(\varkappa)}_l(x)}-\frac{\partial \hat{L}_0(x)}{\partial\hat{\phi}^{(\varkappa)}_l(x)}\right)|0\rangle=m_{\varkappa}\langle0|\hat{\phi}^{(\varkappa)}_l(x)|0\rangle\]
	In any case, if \( m_{\varkappa}\neq0 \), then \( \langle0|\hat{\phi}^{(\varkappa)}_l(x)|0\rangle=0 \). Now we are ready to transform this differential equation into an algebraic one. The idea is to use Poincaré covariance:
	\[\hat{U}(y)\hat{\phi}^{(\varkappa)}_l(x)\hat{U}(-y)=\phi^{(\varkappa)}_l(x+y)\]
	This is an operator-valued function of two arguments x and y. It is a generalized function of the argument x with the parameter y. Then, from the theory of generalized functions, it is easy to show that:
	\[\frac{\partial}{\partial x^{\mu}}\hat{\phi}^{(\varkappa)}_l(x)=(\frac{\partial}{\partial y^{\mu}}\hat{\phi}^{(\varkappa)}_l(x+y))|_{y=0}=i[\hat{P}_{\mu},\hat{\phi}^{(\varkappa)}_l(x)]\]
	We make one more assumption that $\hat{\phi}^{(\varkappa)}_l(x)$ is defined at the point x=0 as an operator on some dense region everywhere in ${\cal H}$, and maps vectors from this region to itself $\hat{\phi}^{(\varkappa)}_l(0)=\hat{\phi}^{(\varkappa)}_l$:
	\[\hat{\phi}^{(\varkappa)}_l(x)=\hat{U}(-x)\hat{\phi}^{(\varkappa)}_l\hat{U}(x)\]
	This assumption is justified because without it it is impossible to talk about the equation for these operators. We can write the equations of motion, for example, for cases where $\varkappa$ are bosons:
	\[\hat{\Pi}(m^2_{\varkappa}\hat{\phi}^{(\varkappa)}_l-\sum_{\mu\nu}\eta^{\mu\nu}[\hat{P}_{\mu},[\hat{P}_{\nu},\hat{\phi}^{(\varkappa)}_l]])\hat{\Pi}=-\hat{\Pi}N\left(\frac{\partial}{\partial \hat{\phi}^{(\varkappa)}_l}V(\hat{\phi}^{(1)}_{l_1},...,\hat{\phi}^{(n)}_{l_n})\right)\hat{\Pi}\]
	Where \( \eta^{\mu\nu} \) is the Euclidean or Minkowski metric. Now we can write the representation of fields on the space $L^{\infty}$:
	\[\hat{\phi}^{(\varkappa)}_l(f_0,f_1(P),...)=(\sum^{+\infty}_{n=1}\int\phi^{(\varkappa)*}_{l,1}(n,P)f_n(P)dx,f_0\phi^{(\varkappa)}_{l,2}(1,P)+\sum^{+\infty}_{n=1}\int\phi^{(\varkappa)}_{l}(1,P;n,P_1)f_n(P_1)dP_1,...)\]
	Using the exact form of the scalar product on $L^{\infty}$, we can find the connection between $\phi^{(\varkappa)}_{l,1}(n,p)$ and $\phi^{(\varkappa)}_{l,2}(n,p)$ if the field is real. But this condition is very difficult, that is why we would solve equations for complex fields (so we don't use some of the conditions on $\hat{\phi}$). Also, we need to show the exact form of the operator $\hat{\Pi}$:
	\[\hat{\Pi}(f_0,f_1(P),...)=(f_0,\sum^{+\infty}_{n=1}\int\Pi(1,P;n,P')f_n(P')dP',..)\]
	Using the fact that $\hat{\Pi}$ commutes with $\hat{P}_{\mu}$, we can obtain a more convenient representation of $\hat{\Pi}$:
	\[\Pi(n,P;n',P')=\Pi(n,n';P)\delta(P-P')\]
	Now we have only one problem—the exact form of the scalar product on the space $L^{\infty}$ is difficult to use. But for the calculations of the Wightman functions, we don't even need to know the explicit form of the scalar product; it's enough that $(1,0,0...)$ is orthogonal to $(0,f_1(p),...)$ for any $f_n(p)$ and that the norm of $(1,0,0...)$ is 1\footnote{Here is the only one moment in the whole article where we used the mathematical assumption about scalar product}. Wightman functions can be expressed through these operators:
	\[\langle0|\hat{\phi}^{(\varkappa_1)}_{l_1}(x_1)...\hat{\phi}^{(\varkappa_n)}_{l_n}(x_{n})|0\rangle=W^{(\varkappa_1\varkappa_2)}_{l_1l_2}(x_1,x_2)\langle0|\hat{\phi}^{(\varkappa_3)}_{l_3}(x_3)...\hat{\phi}^{(\varkappa_n)}_{l_n}(x_{n})|0\rangle+\]
	\[+W^{(\varkappa_1\varkappa_2\varkappa_3)}_{l_1l_2l_3}(x_1,x_2,x_3)\langle0|\hat{\phi}^{(\varkappa_4)}_{l_4}(x_4)...\hat{\phi}^{(\varkappa_n)}_{l_n}(x_{n})|0\rangle+...+W^{(\varkappa_1...\varkappa_n)}_{l_1...l_n}(x_1,x_2,...,x_{n})\]
	Where \( W \) is:
	\[W^{(\varkappa_1...\varkappa_k)}_{l_1...l_k}(x_1,...,x_k)=\sum_{n_1,..n_{k-1}}\int...\int\phi^{(\varkappa_1)*}_{l_1,1}(n_1,P_1)e^{i(P_1(x_1-x_2))}\phi^{(\varkappa_2)}_{l_2}(n_1,P_1;n_2,P_2)...e^{i(P_{k-2}(x_{k-2}-x_{k-1}))}\]
	\[\phi^{(\varkappa_{k-1})}_{l_{k-1}}(n_{k-2},P_{k-2};n_{k-1},P_{k-1})e^{i(P_{k-1}(x_{k-1}-x_{k}))}\phi^{(\varkappa_k)}_{l_k,2}(n_{k-1},P_{k-1})dP_1...dP_{k-1}\]    
	From this point onward, we will focus on \( \phi^3 \) theory with one scalar field. The last step before writing down the equations of motion is the definition of the \( N \)-ordered product. It can be defined by analogy with the free field. The \( N \)-ordered product for free fields can be expressed through vacuum expectations using Wick's theorem, for example:
	\[\hat{\phi}(x_1)....\hat{\phi}(x_n)=N(\hat{\phi}(x_1)....\hat{\phi}(x_n))+\sum^{i=n}_{i=1}\langle0|\hat{\phi}(x_i)|0\rangle N(\hat{\phi}(x_1)...\hat{\phi}(x_{i-1})\hat{\phi}(x_{i+1})...\hat{\phi}(x_n))+...+\langle0|\hat{\phi}(x_1)....\hat{\phi}(x_n)|0\rangle\]
	For interacting fields, the \( N \)-ordered product can be introduced in the same way (as done, for example, in [3]):
	\[m^2\hat{\phi}-\sum_{\mu\nu}\eta^{\mu\nu}(\hat{P}_{\mu}\hat{P}_{\nu}\hat{\phi}+\hat{\phi}\hat{P}_{\mu}\hat{P}_{\nu}-2\hat{P}_{\mu}\hat{\phi}\hat{P}_{\nu})=\lambda(\hat{\phi}^2-\langle0|\hat{\phi}^2|0\rangle\hat{\Pi})\]
	Now we can rewrite the equations for these functions:
	\[(m^2-P_1^2)\phi^*_1(n_1,P_1)=\lambda\sum^{+\infty}_{n_2=1}\int\phi^*_1(n_2,P_2)\phi(n_2,P_2;n_1,P_1)dP_2\]
	\[(m^2-P_1^2)\phi_2(n_1,P_1)=\lambda\sum^{+\infty}_{n_2=1}\int\phi(n_1,P_1;n_2,P_2)\phi_2(n_2,P_2)dP_2\]
	\[(m^2-(P_1-P_2)^2)\phi(n_1,P_1;n_2,P_2)=\lambda\sum^{+\infty}_{n_3=1}\int\phi(n_1,P_1;n_3,P_3)\phi(n_3,P_3;n_2,P_2)dP_3+\]
	\[+\lambda(\phi_2(n_1,P_1)\phi^*_1(n_2,P_2)-\Pi(n_1,n_2;P_1)\delta(P_1-P_2)\sum^{+\infty}_{n_3=1}\int\phi^*_1(n_3,P_3)\phi_2(n_3,P_3)dP_3)\]
	Where $P^2=\sum_{\mu\nu}\eta^{\mu\nu}P_{\mu}P_{\nu}$.
	\subsection{Equations on the separable space $DL$}
	We can do the same as previously but for the space $DL$:
	\[\hat{\phi}^{(\varkappa)}_l(f_0,f(p))=(\int\phi^{(\varkappa)*}_{l,1}(q)f(q)dq,f_0\phi^{(\varkappa)}_{l,2}(p)+\int\phi^{(\varkappa)}_{l}(p,q)f(q)dq)\]
	As before, we can find the Wightman functions without knowing the exact formula for the scalar product:
	\[W^{(\varkappa_1...\varkappa_k)}_{l_1...l_k}(x_1,...,x_k)=\int...\int\phi^{(\varkappa_1)*}_{l_1,1}(p_1)e^{i(p_1(x_1-x_2))}\phi^{(\varkappa_2)}_{l_2}(p_1;p_2)...e^{i(p_{k-2}(x_{k-2}-x_{k-1}))}\]
	\[\phi^{(\varkappa_{k-1})}_{l_{k-1}}(p_{k-2};p_{k-1})e^{i(p_{k-1}(x_{k-1}-x_{k}))}\phi^{(\varkappa_k)}_{l_k,2}(p_{k-1})dp_1...dp_{k-1}\]    
	Where $(p_{i}(x_i-x_{i+1}))=\sum_{\mu\nu}(x^{\mu}_i-x^{\mu}_{i+1})p_i^{\nu}\eta_{\mu\nu}$. Also, we can write the equations of motion for $\phi^3$ theory:
	\[(m^2-\sum_{\mu\nu}\eta^{\mu\nu} tg(p^{\mu}-\frac{\pi}{2}) tg(p^{\nu}-\frac{\pi}{2}))\phi^*_1(p)=\lambda\int\phi^*_1(q)\phi(q,p)dy\]
	\[(m^2-\sum_{\mu\nu}\eta^{\mu\nu}tg(p^{\mu}-\frac{\pi}{2})tg(p^{\nu}-\frac{\pi}{2}))\phi_2(p)=\lambda\int\phi(p,q)\phi_2(q)dq\]
	\[(m^2-\sum_{\mu\nu}\eta^{\mu\nu}(tg(p^{\mu}-\frac{\pi}{2})-tg(q^{\mu}-\frac{\pi}{2}))(tg(p^{\nu}-\frac{\pi}{2})-tg(q^{\nu}-\frac{\pi}{2})))\phi(p,q)=\]
	\[=\lambda(\int\phi(p,r)\phi(r,q)dr+\phi_2(p)\phi^*_1(q)-\Pi(p,q)\int\phi^*_1(r)\phi_2(r)dr)\]
	\section{Some possible solutions}
	The main purpose of this section is to determine whether any possible solutions exist for non-renormalizable theories. First, we introduce a new equation:
	\[N(V(\hat{\phi}(\vec{x}))) - \triangle\hat{\phi}(\vec{x}) + m^2\hat{\phi}(\vec{x}) + \alpha\hat{J}(\vec{x}) = 0\]
	We work in Euclidean $d$-dimensional space with one scalar field $\hat{\phi}(x)$ of mass $m$ (we have done the Wick's rotation). We can build the pertrubation theory for $V$ and $\alpha$. Here we introduce the quantum source $\hat{J}(\vec{x})$, an externally defined operator on $\mathcal{H}$ with the same transformation properties as $\hat{\phi}(\vec{x})$ (translations and rotations). We require that its vacuum expectation value vanishes, $\langle0|\hat{J}(\vec{x})|0\rangle=0$, while the vacuum is not an eigenvector with eigenvalue zero, $\hat{J}(\vec{x})|0\rangle\neq0$. Additionally, we demand that this operator has finite norm. We will provide more details about this operator later.
	
	If we find a solution, then $\mathcal{H}$ can be constructed as the convex hull of vectors $\hat{\phi}(x_1)...\hat{\phi}(x_n)|0\rangle$, which will also be the space where the operator $\hat{J}(x)$ is defined. The space $\mathcal{H}$ is non-trivial because $\hat{J}(\vec{x})|0\rangle\neq0$. It must be infinite-dimensional since $\hat{J}(\vec{x})|0\rangle\neq0$ and because there are no finite-dimensional unitary representations of the Euclidean group with non-trivial realizations of the translation group. We can rewrite this equation in the space $KL_{E}$ using the assumption of existence of $\hat{\phi}(0)$:
	\[N(V(\hat{\phi}))+\hat{\vec{P}}^2\hat{\phi} + \hat{\phi}\hat{\vec{P}}^2 - 2\hat{\vec{P}}\hat{\phi}\hat{\vec{P}} - m^2\hat{\phi} + \mu\hat{\phi} + \alpha\hat{J} = 0\]
	Here the operator $\hat{\Pi}$ is absent because we work in $\mathcal{H}$. For any suitable $\hat{J}$, we can easily find a non-trivial solution for any $\alpha$ using the Banach fixed-point theorem.
	
	We introduce $\mathcal{B}$, the C*-algebra of bounded operators on the Hilbert space $\mathcal{H}$, and define $T(\cdot)$, a linear map on $\mathcal{B}$:
	\[T(\hat{\phi}) = \hat{\vec{P}}^2\hat{\phi} + \hat{\phi}\hat{\vec{P}}^2 - 2\hat{\vec{P}}\hat{\phi}\hat{\vec{P}} + m^2\hat{\phi}\]
	We then define the map $S(\cdot)$:
	\[S(\hat{\phi}) = \hat{\phi} - T^{-1}\left(N(V(\hat{\phi}))+ \hat{\vec{P}}^2\hat{\phi}+\hat{\phi}\hat{\vec{P}}^2 - 2\hat{\vec{P}}\hat{\phi}\hat{\vec{P}} + m^2\hat{\phi} + \alpha\hat{J}\right) = -T^{-1}\left(N(V(\hat{\phi}))+ \alpha\hat{J}\right)\]
	Our goal is to find the fixed point of this map. This map is well-defined because for any operator $\hat{A}\in\mathcal{B}$, $||T^{-1}(\hat{A})||\leq\frac{||\hat{A}||}{m^2}$. We can determine $V(\phi)=\sum^{n=+\infty}_{n=2}c_n\phi^n$.  We can choose such $|\alpha|\leq \frac{m^2R}{2||\hat{J}||}$ and such $R$ that (we are not writing out the most optimal conditions for analyzing solutions to equations):
	\[\sum^{n=+\infty}_{n=2}|c_n|6^nR^n\leq\frac{m^2R}{2}\]
	Then for every $\hat{\phi}$ satisfying $||\hat{\phi}||\leq R$:
	\[||S(\hat{\phi})||\leq R\]
	\[||S(\hat{\phi}_1)-S(\hat{\phi}_2)|| \leq \frac{||\hat{\phi}_1-\hat{\phi}_2||}{2}\]
	Moreover, if $[\hat{U}(1,R),\hat{\phi}]=0$, then $[\hat{U}(1,R),S(\hat{\phi})]=0$. Therefore, we can define the set $\Omega=\{||\hat{\phi}||\leq R; [\hat{U}(1,R),\hat{\phi}]=0\}$. Thus $S(\cdot):\Omega\rightarrow\Omega$, and we can apply the Banach fixed-point theorem to find a non-trivial solution in $\Omega$.
	
	We must highlight two points. First one is that the solution won't sutisfy Hadamard's condition\footnote{$|\langle0|\hat{\phi}(\vec{x})\hat{\phi}(y)|0\rangle|\leq ||\hat{\phi}||^2$ so it will not diverge at coinciding points just like the free Wightman function}. Second one is that there exist such result that Wightman's fields couldn't be bounded operators ([4] exercise 8.8). But in our example we have the external current, wich depends of the coordinate (because $\hat{J}(\vec{x})=\hat{U}(\vec{x},1)\hat{J}\hat{U}^{-1}(\vec{x},1)\neq \hat{J}$). From a physics point of view, the presence of a time-dependent source indicates the pumping of energy into the system (or, conversely, the pumping out), thus the condition on the spectrum of the $\hat{P}^{\mu}$ operator won't be satisfied. From the mathematical point of view we can state that there is no proof of the opposite fact (that in such theories spectrum of the $\hat{P}^{\mu}$ should be in the future light cone). So anyway we don't get the contradiction with the fact that fields couldn't be bounded operatpors.
	\subsection{Renormalization and physical realization}
	We can choose $\hat{J}(\vec{x})$ from the field algebra of a new field (representing a new type of particles). Thus we have two scalar fields: $\hat{\phi}(\vec{x})$ and $\hat{\varphi}(\vec{x})$. The operator $\hat{J}(\vec{x})$ is an element of the operator algebra of $\hat{\varphi}(\vec{x})$. 
	
	In QED, we could define an external electromagnetic field by some function. Here we consider a quantum theory where an external field is represented not by a function (c-number) but by an operator as if the external field were quantum rather than classical. The same situation occurs here but for scalar fields. 
	
	We can choose a specific form for the external source, such as $\hat{J}(\vec{x})=sin(\beta\hat{\varphi}(\vec{x}))$. We do not imply that there exists another field interacting with $\hat{\phi}$ through this source. Rather, we mean that this operator resembles $sin(\beta\hat{\varphi}(\vec{x}))$, where $\hat{\phi}$ is an abstract free field. Then we can use diagram techniques to find all Wightman functions of the operators $\hat{\phi}$. The new particles would be similar to ghosts in QCD, existing only in loops.
	
	However, we can view this model from another perspective. We can consider two fields $\hat{\phi}$ and $\hat{\varphi}$:
	\[N(V(\hat{\phi}(\vec{x})))-\triangle\hat{\phi}(\vec{x})+m^2\hat{\phi}(\vec{x})+\alpha sin(\beta\hat{\varphi}(\vec{x}))= 0\]
	\[-\triangle\hat{\varphi}(\vec{x})+M^2\hat{\varphi}(\vec{x}) + \alpha\hat{\phi}(\vec{x})cos(\beta\varphi(\vec{x})) = 0\]
	
	If we take $M\rightarrow+\infty$, the second equation can be solved with very high accuracy by a free field in this limit. This demonstrates how such a non-physical entity as a quantum source can appear in the equations of motion.
	
	In any case, it is clear that this theory could be non-renormalizable, we have previously proved that non-trivial solutions to the equations of motion exist for any $V$ (for any analytical function $V$). This shows that renormalizability is not a necessary condition for a physical theory.
	\section*{Conclusions}
	In the fourth section, we have managed to reduce the equation of motion for \(\phi^3\) to a quadratic matrix equation with matrices over a separable Hilbert space (with a comfortable spectrum of $\hat{P}_{\mu}$). The solution of these equations is the first open and very important question for physics. In the fifth section, we proved the existence of solutions even in case when the theory is non-renormalizable. But also we introduced a mathematical problem of generalization of the classification of all solutions, because we are not sure that the only one solution that we found is really that one that could be realized in physics (especially considering that it does not satisfy Hadamard's condition). Morover there is still open problem if there exist any solution of $\phi$-cubed theory or any other more familiar theories.
	\section*{Acknowledgments}
	I would like to express my gratitude to E. T. Akhmedov for supporting this work and to M.A. Pliev for explaining to me the need to clearly state the mathematical assumption, as it could not be proved.
	\appendix 
	\section{Equations of Motion for QCD}
	In QCD (d=4), we have $N$ fermions $\psi^{i}_{j}(x)$—quarks, where $i=1,..,N$ is a color index and $j=1,..4$ are components of Dirac's bispinor. Also, we have $\gamma^{\mu}$—Dirac's gamma matrices. Additionally, there are $N^2-1$ types of gluons—massless particles with four components $G_{a}^{\mu}(x)$, where $a=1,...,N^2-1$. There are also $t^{a}_{k,k'}$—generators of the SU(N) group, and $f^{abc}$—structure constants. We will rewrite the equations as in the fifth section:
	\[\gamma^{\mu}_{i,i'}[\hat{P}_{\mu},\hat{\psi}^{k}_{i'}]+m\hat{\psi}^{k}_{i}+g_s\gamma^{\mu}_{i,i'}t^{a}_{k,k'}(\hat{G}^{a}_{\mu}\hat{\psi}^{k'}_{i'}-\hat{\Pi}\langle0|\hat{G}^{a}_{\mu}\hat{\psi}^{k'}_{i'}|0\rangle)=0\]
	\[[\hat{P}^{\nu},[\hat{P}_{\nu},\hat{G}^{a}_{\mu}]]-[\hat{P}_{\mu},[\hat{P}^{\nu},\hat{G}^{a}_{\nu}]]-ig_sf^{abc}([\hat{P}^{\nu},\hat{G}^{b}_{\nu}\hat{G}^{c}_{\mu}]+\hat{G}^{b}_{\nu}([\hat{P}^{\nu},\hat{G}^{c}_{\mu}]+[\hat{P}_{\mu},\hat{G}^{c \nu}])-\hat{\Pi}\langle0|\hat{G}^{b}_{\nu}(\hat{P}^{\nu}\hat{G}^{c}_{\mu}+\hat{P}_{\mu}\hat{G}^{c \nu})|0\rangle)-\]
	\[-g_s^2f^{abc}f^{cde}(\hat{G}^{b}_{\nu}\hat{G}^{d \nu}\hat{G}^{e\mu}-\hat{G}^{b}_{\nu}\langle0|\hat{G}^{d \nu}\hat{G}^{e\mu}|0\rangle-\hat{G}^{e\mu}\langle0|\hat{G}^{b}_{\nu}\hat{G}^{d \nu}|0\rangle-\hat{G}^{d \nu}\langle0|\hat{G}^{b}_{\nu}\hat{G}^{e\mu}|0\rangle-\langle0|\hat{G}^{b}_{\nu}\hat{G}^{d \nu}\hat{G}^{e\mu}|0\rangle\hat{\Pi})+\]
	\[+g_s\hat{\psi}^{k\dagger}_{i}\gamma^{0}_{i,i'}\gamma_{\mu i',i''}t^{a}_{k,k'}\hat{\psi}^{k'}_{i''}-g_s\langle0|\hat{\psi}^{k\dagger}_{i}\gamma^{0}_{i,i'}\gamma_{\mu i',i''}t^{a}_{k,k'}\hat{\psi}^{k'}_{i''}|0\rangle\hat{\Pi}=0\]
	Where $\langle0|\hat{\psi}|0\rangle=0$ as in the fourth section, $m\neq0$. But gluons are massless, so the same equality holds for them by physical meaning (we are interested in the potential between two quarks, so $\langle0|\hat{G}^{a}_{\mu}|0\rangle=0$). We can also rewrite these equations on the spaces $DL$ and $L^{\infty}$. 
	\bibliography{literature}
	
	[1] A. S. Wightman \textit{Quantum field theory in terms of vacuum expectation values. - Phys.Rev., v.101, p. 860.}
	
	[2] Brünning E. \textit{On the characterization of relativistic quantum field theories in terms of finitely many vacuum expectation values, I, II.- Commun. Math. Phys., v. 58, p. 139, 166.}
	
	[3] A.S. Wightman, \textit{Problems in Relativistic Dynamics of Quantum Fields}.
	
	[4] N.N. Bogolyubov, A.A. Logunov, A.I. Oksak, I.T. Todorov, \textit{General Principles of Quantum Field Theory}.
	
	[5] N.I. Akhiezer, I.M. Glazman, \textit{Theory of Linear Operators in Hilbert Space}.

	\bibliographystyle{unsrt}
\end{document}